\documentclass[]{aa}  
\usepackage{graphicx}
\usepackage{txfonts}
\usepackage{lipsum}
\usepackage[dvipsnames]{xcolor}
\usepackage{stfloats}
\usepackage{multirow}
\usepackage[version=4]{mhchem}
\usepackage[utf8]{inputenc}
\usepackage{booktabs}
\usepackage{amssymb}
\usepackage{xspace}

\usepackage{array}
\newcolumntype{L}[1]{>{\raggedright\let\newline\\\arraybackslash\hspace{0pt}}m{#1}}
\newcolumntype{C}[1]{>{\centering\let\newline\\\arraybackslash\hspace{0pt}}m{#1}}
\newcolumntype{R}[1]{>{\raggedleft\let\newline\\\arraybackslash\hspace{0pt}}m{#1}}

\newcommand{\micron}{\textmu m\xspace}

\definecolor{darkorange}{rgb}{1.0, 0.55, 0.0}

\usepackage{soul}

\begin{document} 
   \title{The edge-on protoplanetary disk HH~48~NE}
   \subtitle{II. Modeling ices and silicates}
   \titlerunning{HH~48~NE: II. ices and silicates}

   \author{J.A. Sturm\inst{1}\thanks{sturm@strw.leidenuniv.nl}
          \and
          M.K. McClure\inst{1}
          \and
          J.B. Bergner\inst{2}
          \and
          D. Harsono\inst{3}
          \and
          E. Dartois\inst{4}
          \and
          M.N. Drozdovskaya\inst{5}
          \and
          S. Ioppolo\inst{6}
          \and
          K.I. \"Oberg\inst{7}
          \and
          C.J. Law\inst{7}
          \and
          M.E. Palumbo\inst{8}
          \and
          Y.J. Pendleton\inst{9}
          \and
          W.R.M. Rocha\inst{1,10}
          \and
          H. Terada\inst{11,12}
          \and
          R.G. Urso\inst{8}
          }

   \institute{Leiden Observatory, Leiden University, P.O. Box 9513, NL-2300 RA Leiden, The Netherlands
    \and
    Department of Chemistry, University of California, Berkeley, California 94720-1460, United States
    \and
    Institute of Astronomy and Astrophysics, Academia Sinica, No. 1, Sec. 4, Roosevelt Road, Taipei 10617, Taiwan, R. O. C.
    \and
    Institut des Sciences Mol\'eculaires d’Orsay, CNRS, Univ. Paris-Saclay, 91405 Orsay, France
    \and 
    Center for Space and Habitability, Universit\"at Bern, Gesellschaftsstrasse 6, 3012, Bern, Switzerland 
    \and
    Center for Interstellar Catalysis, Department of Physics and Astronomy, Aarhus University, Ny Munkegade 120, Aarhus C 8000, Denmark
    \and 
    Center for Astrophysics \textbar\ Harvard \& Smithsonian, 60 Garden St., Cambridge, MA 02138, USA
    \and
    INAF – Osservatorio Astrofisico di Catania, via Santa Sofia 78, 95123 Catania, Italy
    \and 
    Department of Physics, University of Central Florida, Orlando, FL 32816, USA 
    \and
    Laboratory for Astrophysics, Leiden Observatory, Leiden University, PO Box 9513, 2300 RA Leiden, The Netherlands
    \and
    TMT International Observatory, 100 W Walnut St, Suite 300, Pasadena, CA USA
    \and
    National Astronomical Observatory of Japan, National Institutes of Natural Sciences (NINS), 2-21-1 Osawa, Mitaka, Tokyo 181-8588, Japan
    }

    
    \date{Received XXX; accepted YYY}

    \abstract
   {The abundance and distribution of ice in protoplanetary disks is critical to understand the linkage between the composition of circumstellar matter and the composition of exoplanets. 
   Edge-on protoplanetary disks are a useful tool to constrain such ice composition and its location in the disk, as ice spectral signatures can be observed in absorption against the continuum emission arising from the warmer central disk regions.}
   {The aim of this work is to model ice absorption features in protoplanetary disks and determine how well the abundance of the main ice species across the disk can be determined within the uncertainty of the physical parameter space. 
   The edge-on protoplanetary disk around HH~48~NE, a target of the \textit{James Webb} Space Telescope Early Release program IceAge, is used as a reference system.}
   {We use the full anisotropic scattering capabilities of the radiative transfer code \texttt{RADMC-3D}, to raytrace the mid-infrared continuum. Using a constant parameterized ice abundance, ice opacities are added to the dust opacity in regions wherever the disk is cold enough for the main carbon, oxygen and nitrogen carriers to freeze out.}
   {The global abundance, relative to the dust content, of the main ice carriers in HH~48~NE can be determined within a factor of 3, when taking the uncertainty of the physical parameters into account.
   Ice features in protoplanetary disks can be saturated at an optical depth of $\lesssim$1, due to local saturation. 
    Ices are observed at various heights in the disk model, but in this model spatial information is lost for features at wavelengths >7~\micron when observing with \textit{James Webb} Space Telescope, due to the lower angular resolution at longer wavelengths. 
   Spatially observed ice optical depths cannot be directly related to column densities, as would be the case for direct absorption against a bright continuum source, due to radiative transfer effects.
    Vertical snowlines will not be a clear transition due to the radially increasing height of the snowsurface, but their location may be constrained from observations using radiative transfer modeling. Radial snowlines are not really accesible.
   Not only the ice abundance, but also inclination, settling, grain size distribution and disk mass have strong impact on the observed ice absorption features in disks.
   Relative changes in ice abundance can be inferred from observations only if the source structure is well constrained.
}
   {}

    \maketitle
\section{Introduction}
Ices constitute the bulk reservoir of volatiles like C, N, and O in cold protoplanetary disks \citep{pontoppidan2014}, and are thought to be the carriers and precursors of astrophysically complex organic molecules \citep{boogert2015}. 
Ices on dust grains and the macro-molecular residues formed from their processing in space environments are often recognized as important carriers of organic compounds of astrobiological relevance \citep[e.g.,][]{Baratta2019,johansen2021,sossi2022}. 
It is crucial to investigate the processing of small molecules, like CO, \ce{CO2}, and the origin of the diversity of complex organics in planet-forming regions in protoplanetary disks to understand the origin of chemical complexity on planets.

Previously, the thermal snow-surfaces of the molecules CO, CS and \ce{H2O} have been traced directly in protoplanetary disks using observations of their gas emission lines \citep{Qi2013_CO_snowline,pinte2018,Vanthoff2020,Podio2020}.
Also, ices have been traced indirectly in studies focusing on radial transport of volatiles frozen out on large dust grains in the disk midplane \citep{Zhang2020drift,Sturm2022CI}, and through their physical imprint on the grain properties \citep{Banzatti2015,Cieza2016}.
\ce{H2O} ice is observed directly towards multiple class II disks \citep[e.g.][]{terada2007,aikawa2012,terada2012,McClure2015,Terada2017} using ground-based observations of Infrared
Camera and Spectrograph (IRCS) on the Subaru telescope and the AKARI and Herschel space telescopes.
However, until the advent of the recently launched \textit{James Webb} Space Telescope (JWST), direct observations  of ices other than \ce{H2O} in disks were hardly possible since Earth's atmosphere absorbs radiation at those wavelengths efficiently and previous space-based instruments lacked enough sensitivity and spatial resolution.
\ce{CO2}, and tentatively CO, ice features have been detected previously in a handful of edge-on Class II disks with the AKARI space telescope \citep{aikawa2012}.
Other ices have been observed towards multiple young Class 0/I protostars with edge-on disks (e.g.,  \citealt{Pontoppidan2005} and \citealt{aikawa2012})  but it is hard to decipher what fraction of ices the disk in those systems actually contributes to the absorption bands, as the ice absorption can be dominated by the envelope and foreground cloud.

JWST Near Infrared Spectrograph \citep[ NIRSpec, 0.6 - 5.3~\micron][]{Jakobsen2022_nirspec} and Mid-InfraRed Instrument \citep[MIRI, 4.9-28.1~\micron;][]{Rieke2015_miri} observations of later stage edge-on disks (>2 Myr), where the envelope has long dissipated, will shed light on the composition of ices in protoplanetary disks and their spatial distribution in the disk.
Space-based JWST is sensitive enough to observe many more ice species in protoplanetary disks, and has the resolving capability to constrain absorption strengths spatially across the disk.
However, detailed modeling of these edge-on disks is necessary to interpret what the observations reveal about the physical and chemical structure of such disks.
Ices in edge-on protoplanetary disks are revealed in absorption against the inner disk continuum and/or the scattered light. 
Both stellar photons and photons emitted from warm dust in the (<5~au) inner regions of the disk are emitted/scattered off small dust grains on the surface of the outer disk, and get absorbed at specific wavelengths by ices located in the cold regions in the disk (see Appendix \ref{app:comparison_hyperion} for a detailed breakdown of the observed continuum emission).
For a sketch of the basic principles, see Fig. 3 in \citet{terada2007}.

\citet{Ballering2021} used advanced chemical modeling to show that the major ice species could likely be detected with JWST. 
That work focuses on the observability of chemical differences with the current generation of observatories and finds that the strength of the absorption features could reveal if the ices are inherited from the protostellar phase or are formed after a molecular reset upon entering the disk.
\citet{Arabhavi2022} implemented ice opacities in Protoplanetary Disk Model (ProDiMo) and looked at the impact of the chemistry and ice distribution across dust grains on the spectral ice features in the JWST wavelength range. 
\citet{dartois2022} modelled the influence of grain growth on the ice features spectroscopic profiles in both dense clouds and disks observations.
In this work, we extend the modeling analysis of ices in protoplanetary disks and focus on how the physical disk parameters affect the JWST spectra in edge-on protoplanetary disks, and quantify how robustly we can constrain ice abundances. 

While we apply our modeling to one specific disk, HH~48~NE, this work can readily be generalized to all edge-on disks. 
HH~48~NE is part of the JWST early release science program Ice Age (proposal ID: 1309, PI: McClure), which studies the evolution of ices during star formation from dark clouds \citep{McClure2023} till protoplanetary disks.
The disk geometry and stellar parameters are explored in Paper I of this paper series \citep{PaperI}. 
HH~48~NE is a nearly edge-on disk first observed with Hubble space telescope (HST) \citep[][]{Stapelfeldt2014} and subsequently with Atacama large (sub-)millimeter array (ALMA) \citep[][]{Villenave2020}. 
In Paper I, we showed that the 4-5 Myr old star is of spectral type K7, with a luminosity of 0.4 L$_\odot$. The 200~au scattered light disk is inclined at 82.3\degr, and likely has a 55~au cavity where dust is depleted by 2 orders of magnitude. 

In this work, we use a simple, parameterized abundance structure for the main ice species \ce{H2O}, \ce{CO}, \ce{CO2}, \ce{NH3}, \ce{CH4}, and \ce{CH3OH} to reveal the disk physical parameters required to assess the ice abundances and chemistry, and how accurately we can constrain the chemistry in HH~48~NE given the uncertainty in disk geometry.
The structure of this paper is as follows: we first describe the model setup that we used in Sect. \ref{sec:model_description}. In Sect. \ref{sec:results}, we present the ice spectrum in the fiducial model and the sensitivity of the ice features to various parameters. In Sect. \ref{sec:discussion}, we discuss the implications of these results in the analysis of HH~48~NE and edge-on disks in general. 
Sect. \ref{sec:conclusion} summarizes the results and gives final conclusions.

\section{Modeling}
\label{sec:model_description}
\subsection{Continuum model}
Our modeling is based on the full anisotropic scattering radiative transfer capabilities of \texttt{RADMC-3D}  \citep{Dullemond2012radmc3d}. In the first stage the temperature is determined, after which we distribute the ices in cold disk regions and ray-trace the model to simulate observations. 
The specific steps of our modeling procedure are described in the following sections. 
All parameters are adopted from the best fitting model in Paper I, that reproduces the spectral energy distribution well, and listed in Table \ref{tab:fid_mod_props}.

\subsubsection{Model setup}
The model setup that we used is fully parameterized, assuming an azimuthally symmetric disk with a power law density structure and an exponential outer taper \citep{lynden-bell1974}
\begin{equation}
\label{eq:surface_density_profile}
\Sigma_{\mathrm{dust}}=\frac{\Sigma_{\mathrm{c}}}{\epsilon} \left(\frac{r}{R_{\mathrm{c}}}\right)^{-\gamma} \exp \left[-\left(\frac{r}{R_{\mathrm{c}}}\right)^{2-\gamma}\right],
\end{equation}
where $r$ is the radial distance to the star in the plane of the disk in au, $\Sigma_\mathrm{c}$ is the surface density at the characteristic radius $R_\mathrm{c}$ in g~cm$^{-2}$, $\gamma$ the power law index and $\epsilon$ the gas-to-dust ratio.

\begin{figure*}
    \centering
    \includegraphics[width = \textwidth]{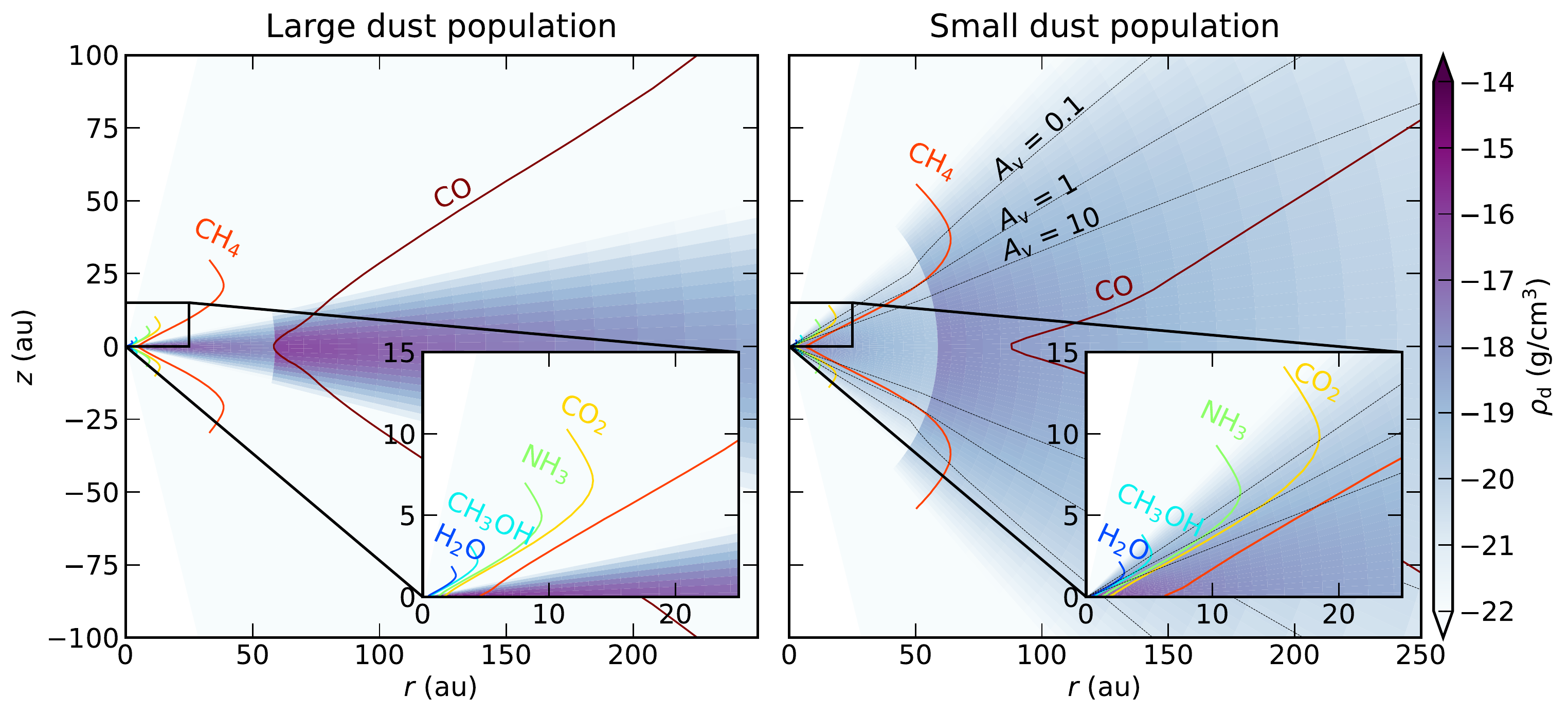}
    \caption{Dust density setup in the model of the large dust population (left) and small dust population (right). The colored contours follow the dust temperature at which the considered molecules freeze out according to their desorption temperature given in Table \ref{tab:iceprops}. The black dotted lines denote a visual extinction ($A_\mathrm{v}$) of 0.1, 1 and 10~mag as measured from the stellar position. The onset of ice formation is set to $A_\mathrm{v}$ < 1.5~mag in the fiducial model.}
    \label{fig:denstemp}
\end{figure*}

The height of the disk is described by 
\begin{equation}
\label{eq:disk_height}
h=h_{\mathrm{c}}\left(\frac{r}{R_{\mathrm{c}}}\right)^{\psi},
\end{equation}
where $h$ is the aspect ratio, $h_{\mathrm{c}}$ the aspect ratio at the characteristic radius and $\psi$ the flaring index.
The dust has a vertical Gaussian distribution 
\begin{equation}
    \label{eq:small_grains_distribution}
    \rho_{\mathrm{d}}=\frac{\Sigma_{\mathrm{dust}}}{\sqrt{2 \pi} r h} \exp \left[-\frac{1}{2}\left(\frac{\pi / 2-\theta}{h}\right)^{2}\right],
\end{equation}
where $\theta$ is the opening angle from the midplane as seen from the central star.
The dust sublimation radius is determined using $r_{\rm in}$ =  0.07$\sqrt{L/L_{\odot}}$, assuming a dust sublimation temperature of 1600 K, which corresponds to 0.045~au.

\begin{table}[!t]
    \centering
    \caption{Properties of the fiducial model constrained by extensive MCMC fitting using the spectral energy distribution and resolved HST and ALMA observations in Paper I.}
    \label{tab:fid_mod_props}
    \begin{tabular}{L{.45\linewidth}L{.45\linewidth}}
        \bottomrule
        \toprule
        parameter (unit)                & value\\
        \midrule
        $L_{\rm s}$ (L$_\odot$)         & 0.41 \\
        $T_{\rm s}$ (K)                 & 4155 \\
        $R_{\rm c}$ (au)                & 87   \\
        $h_{\rm c}$                     & 0.24 \\ 
        $\psi$                          & 0.13 \\
        $i$ $(^{\rm o})$                & 82.3 \\
        $\gamma$                        & 0.81 \\
        $M_{\rm gas}$ (M$_{\odot}$)     & 2.7$\times10^{-3}$ \\
        $f_\ell$                        & 0.89 \\
        X                               & 0.2  \\
        $a_{\rm min}$ (\micron)         & 0.4  \\
        $a_{\rm max,s}$ (\micron)       & 7    \\
        $a_{\rm max,l}$ (\micron)       & 1000 \\
        $R_{\rm cav}$ (au)              & 55   \\
        $\delta_{\rm cav}$              & 1.6$\times10^{-2}$\\
        \bottomrule
    \end{tabular}
\end{table}

Dust settling is parameterized by introducing a second dust population containing large grains, that is limited in height to $Xh$, where $X$ is a constant between 0 and 1.
The mass fraction of the total dust mass that resides in the large grain population is defined as $f_\mathrm{\ell}$. 
We assumed a grain composition consistent with ISM dust, consisting of a mass fraction of 85\% amorphous pyroxene Mg$_{0.8}$Fe$_{0.2}$SiO$_3$, 15\% amorphous carbon and a porosity of 25\%.
The two dust populations follow a powerlaw size distribution with a fixed slope of -3.5.
See e.g. \citet{weingartner2001} and \citet{andrews2011} for observational justification of these parameters.
The grain size limits of the small dust population are $a_{\rm min}$ = 0.4~\micron and $a_{\rm max}$ = 7~\micron, as determined with Markov Chain Monte Carlo (MCMC) fits using the spectral energy distribution (SED) and the resolved HST and ALMA observations in Paper I.
The minimum grain size of the large dust population is the same as the minimum size of the small dust population, the maximum grain size of the large dust population is adopted to be 1~mm. 
Dust and ice opacities were calculated with \texttt{OpTool} \citep{Dominik2021optool}, using the Distribution of Hollow Spheres \citep[DHS;][]{Min2005DHS} approach to account for grain shape effects. 
The phase function is truncated at 3\degr, which means that forward scattering within this range is treated as if the photon package did not have any interaction with the grain at all.

\subsubsection{Dust scattering}
We used the full anisotropic scattering capabilities of \texttt{RADMC-3D}, as isotropic scattering assumptions can have a significant impact on the amount of observed scattered light in the near- to mid-infrared \citep{Pontoppidan2007}. 
Additionally, the direction from which the scattered light originates is important for modeling the ices in the disk. 
The spectral range where most ice absorption features occur is dominated by scattered light that is emitted in the warm inner region of the disk and scattered off in the outer disk (see Fig. \ref{fig:hyperion_comparison}). 
As the ices are localized in the lower layers in the disk, directional dependence is critically important for the amount of light that is absorbed by the ices. 

\begin{table*}[!b]
\caption{Adopted ice properties: Desorption temperature (taken as the recommended peak desorption temperature in \citet{Minissale2022} under high density conditions), Abundance with respect to total H \citep{Ballering2021}, Molecular weight, disk regions in which the ice is distributed and references for the used optical constants. Disk regions are numbered from warm to cold, increasingly adding one ice component.}
\label{tab:iceprops}
\begin{tabular}{L{.18\linewidth}|L{.07\linewidth}L{.09\linewidth}L{.07\linewidth}L{.15\linewidth}L{.3\linewidth}}
\toprule
Opacity   & $T_{\rm des}$  & Abundance  & M$_{n}$  & Deposited in & Reference optical constants\\
component      & (K)            & (ppm)&&disk region &\\
\midrule
Dust  & 1600 & -    & -    & {[}1, 2, 3, 4, 5, 6, 7{]}\\
\ce{H2O}   & 155  & 80   & 18   & {[}2, 3, 4, 5, 6, 7{]} & \citet{Warren2008}  \\
\ce{CH3OH} & 128  & 4.8  & 32   & {[}3, 4, 5, 6, 7{]} & \citet{Gerakines2020}  \\
\ce{NH3}   & 96   & 4.8  & 17   & {[}4, 5, 6, 7{]} & \citet{martonchik1984}    \\
\ce{CO2}   & 80   & 22.4 & 44   & {[}5, 6, 7{]} & \citet{Warren1986}           \\
\ce{CH4}   & 47   & 3.60 & 16   & {[}6, 7{]} & \citet{Gerakines2020}           \\
\ce{CO}    & 20   & 99.2 & 28   & {[}7{]} & \citet{Palumbo2006}                \\
\midrule
\end{tabular}

\end{table*}

\begin{figure*}
    \centering
    \includegraphics[width = \textwidth]{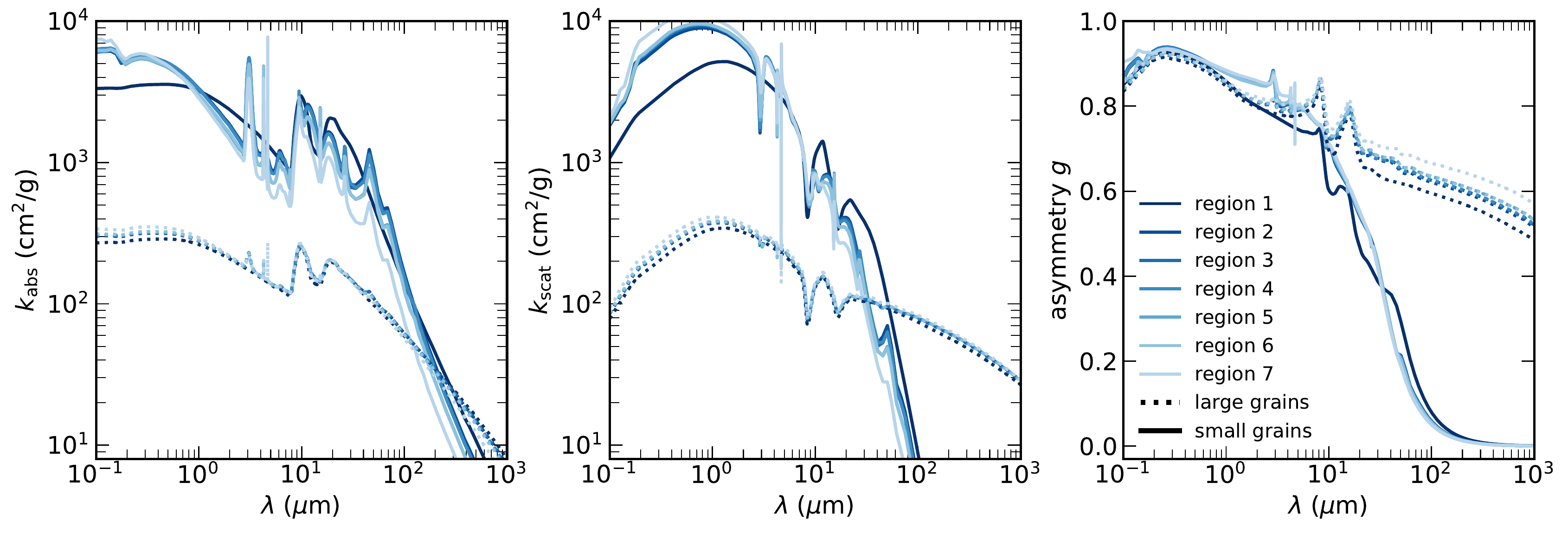}
    \caption{Opacities used for the 7 different regions in the disk. Left: absorption opacity, center: effective scattering opacity, right: asymmetry parameter g. The regions are numbered with decreasing desorption temperature, further specified in Table \ref{tab:iceprops}.}
    \label{fig:opacities}
\end{figure*}

\subsection{Ice distribution}
For evaluation of the ices in the model, we assumed a constant molecular abundance throughout the disk, to separate the effect on the ice features of physical disk parameters from chemistry.
The molecular abundances are adopted from the initial abundances in the inheritance model in \citet{Ballering2021}, based on dark cloud ice surveys \citep{Oberg2011, boogert2015}, for an easy comparison between their and our approach.
All molecules were assumed to be frozen out wherever the dust temperature is below the desorption temperature of that particular species (see Table \ref{tab:iceprops} for details).
The ice was limited vertically by a visual extinction ($A_{\rm v}$) limit from the star, supported by both dark cloud observations \citep{boogert2015} and protoplanetary disk models \citep{visser2018niso,Arabhavi2022}. 
This limit is set to $A_{\rm v}$ = 1.5~mag in the fiducial model, based on the lower limit where ices are observed in dark clouds towards background stars \citep{boogert2015}.
We do not include a background radiation field, as there are no early spectral type stars in the vicinity of HH~48. 
For sources with a strong background radiation field the ice can be photo-desorbed in a significant area of the disk, since this radiation will encounter the cold, translucent parts of the disk first.
Future JWST observations of ices in a larger sample of protoplanetary disks will shed more light on the effect of the external radiation field on ices in disks.

To ray-trace the ice features, time and computational efficiently in \texttt{RADMC-3D}, we divided the disk into 7 regions, one for each considered ice species, progressively adding ices towards cooler regions in the disk based on their respective absorption temperature (see Table \ref{tab:iceprops} and Fig. \ref{fig:denstemp}). The average density of the dust and ice composite is then calculated in every region using

\begin{equation}
\rho_{\text {ice }}=\rho_{\rm gas} \frac{x_{\mathrm{ice}} M_{\mathrm{ice}}}{x_{\rm gas} M_{\rm gas}},
\end{equation}
where $x$ is the abundance and $M$ the mean molecular weight. The gas consists predominantly of \ce{H2} and He, and has an abundance of $x_{\rm gas}$ = 0.64 and a mean molecular weight of 2.44.

The ice is distributed over the small and large grain population according to their surface area, following the description in \citet{Ballering2021}
\begin{equation}
f_{\text {ice }, \ell}=\frac{f_{\text {dust }, \ell}}{\sqrt{a_{\max , \ell} / a_{\max , \mathrm{s}}}\left(1-f_{\mathrm{dust}, \ell}\right)+f_{\mathrm{dust}, \ell}},
\end{equation}
where $a_{\rm max,s}$ is the maximum grain size of the small dust population, $a_{\rm max,\ell}$ is the maximum grain size of the large dust population and $f_{\rm dust,\ell}$ is the mass fraction of grains in the large dust population.
In each of the two dust populations the ice is distributed over the grains assuming a constant core-mantle mass ratio. 
This approach is consistent with dust grains that are a coagulation of small dust grains covered in ice, and for preferred freezeout on large grains due to a lower surface energy barrier \citep{Powell2022}. 
We would like to note that the choice of ice partitioning for large grains does not significantly influence the results as most ice absorption we see happens in the optically thin upper regions of the disk where the large dust population is excluded by settling.
A distribution over the grains assuming a constant mantle thickness in the small dust population would have a significant effect, as shown recently by \citet{Arabhavi2022}.

Table \ref{tab:iceprops} lists the laboratory measured ice optical constants used for the opacity calculation, that are available through \texttt{OpTool}.  
We mixed the optical constants together in each of the regions using the Bruggeman rule \citep{bruggeman1935}, assuming homogeneous mixing of the ice components in an icy mantle on top of the dust core.
We assumed only pure ice components, measured at a singular temperature (see Table \ref{tab:iceprops} for the references), and thus did not take into account how the precise shape of the feature and peak position vary with temperature and the ice matrix.
A more detailed analysis on the shape of the absorption features and the dependence on the chemical environment in protoplanetary disks will be given in a forthcoming paper.

We used a well sampled grid of 200 x 100 x 25 cells in radial, vertical and azimuthal directions, respectively and used 10$^8$ photons ($\tau_{\rm peak}$ $\sim$0.1) to determine the temperature structure in the model.
After deposition of the ices we repeat the temperature calculation to be sure that adding the reflective layer to the opacities does not change the internal temperature in the model. 
In theory, one could do multiple iterations over the ice deposition and temperature calculations, to ensure that the regions where the ice are distributed are accurately determined. 
However, to limit the computing time, we ignore the small deviations in deposition area. 
The temperature differences between different iterations are small and converge rapidly \citep{Arabhavi2022}.

In Fig. \ref{fig:opacities} we present the opacities in the 7 different regions for both the small and large grain population. 
Besides addition of molecular ice bands arising over relatively narrow wavelength ranges, the ice addition modifies the overall shape of the continuum slightly.
The total opacity of the small grains dominates over the large grains at shorter wavelengths, while the large grains take over at longer wavelengths. 

The model is ray-traced at $309$ wavelengths, hand picked to resolve the mid-infrared ice features. 
For ray-tracing we used 10$^7$ photon packages per frame, with image dimensions of 300 pixels in size, and a pixel size of 2~au. 
The 2D images are afterwards convolved with an oversampled instrumental point spread function (PSF) for NIRSpec (<5~\micron) or MIRI (>5~\micron) using the \texttt{Python} package \texttt{WebbPSF} \citep{webbpsf2012}, to create mock observations that can directly be compared with the observed data. 

\begin{figure}[!b]
    \centering
    \includegraphics[width=1\linewidth]{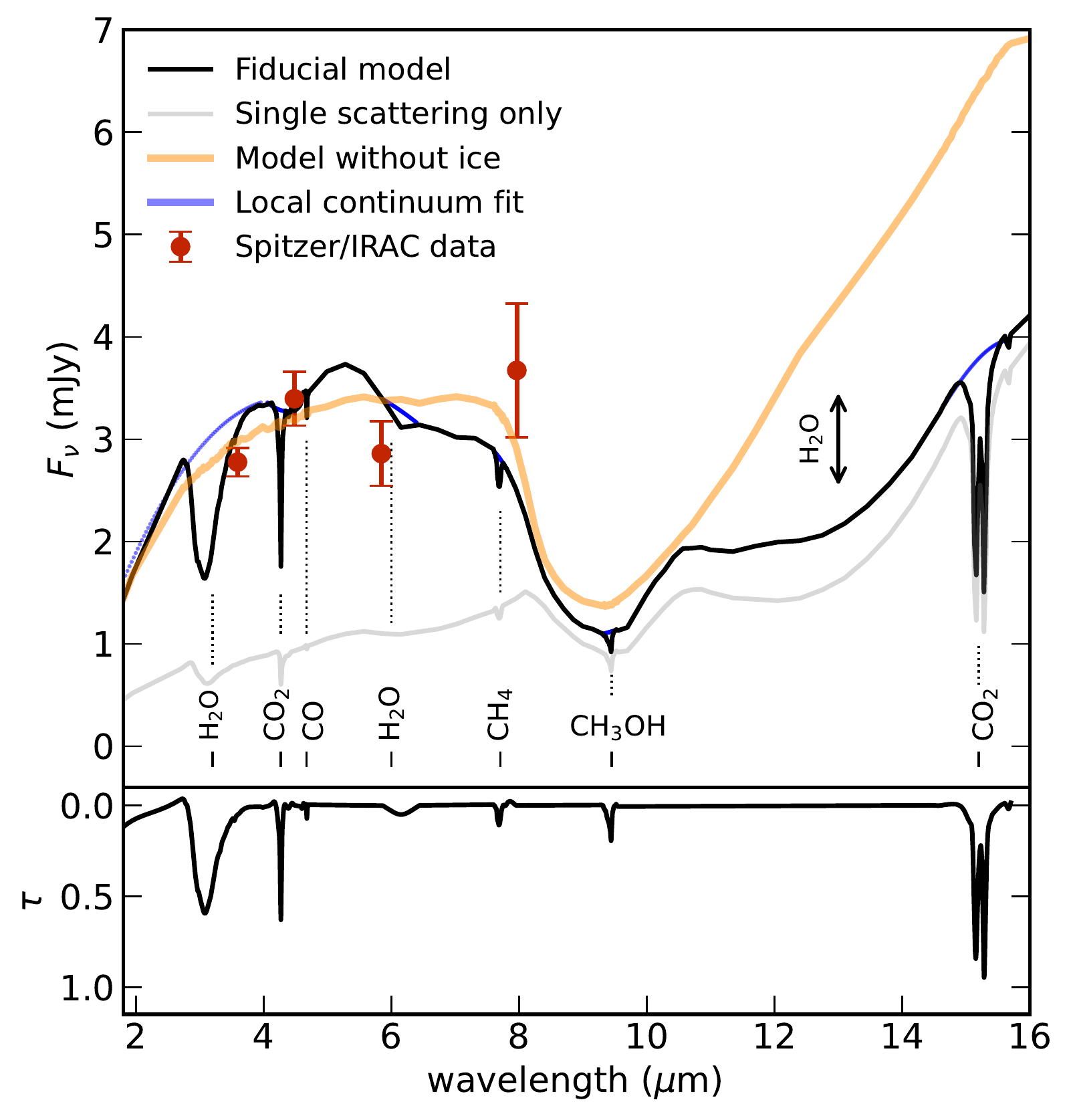}
    \caption{Integrated spectrum of the fiducial model, with photometric data points (red circles) and the best fitting dust-only model (orange line) for reference. The local continuum used to calculate the optical depths of the sharp ice features is shown in blue. The fiducial model excluding double scattering is shown in grey. The lower panel shows the optical depth of the analyzed ice features in the fiducial spectrum.}
    \label{fig:fiducial_ice_sed}
\end{figure}

\section{Results}
\label{sec:results}
Here we discuss the results of the ice modeling. 
We first discuss the specifics of the different ice species in the fiducial model. 
After that we present various grids, changing one physical parameter at a time, to determine the sensitivity of the optical depth of the ice features to the physical parameters and to what extent we can constrain ice abundances spatially in the disk.

\subsection{Fiducial model}
\begin{figure*}
    \centering
    \includegraphics[width = 1\textwidth]{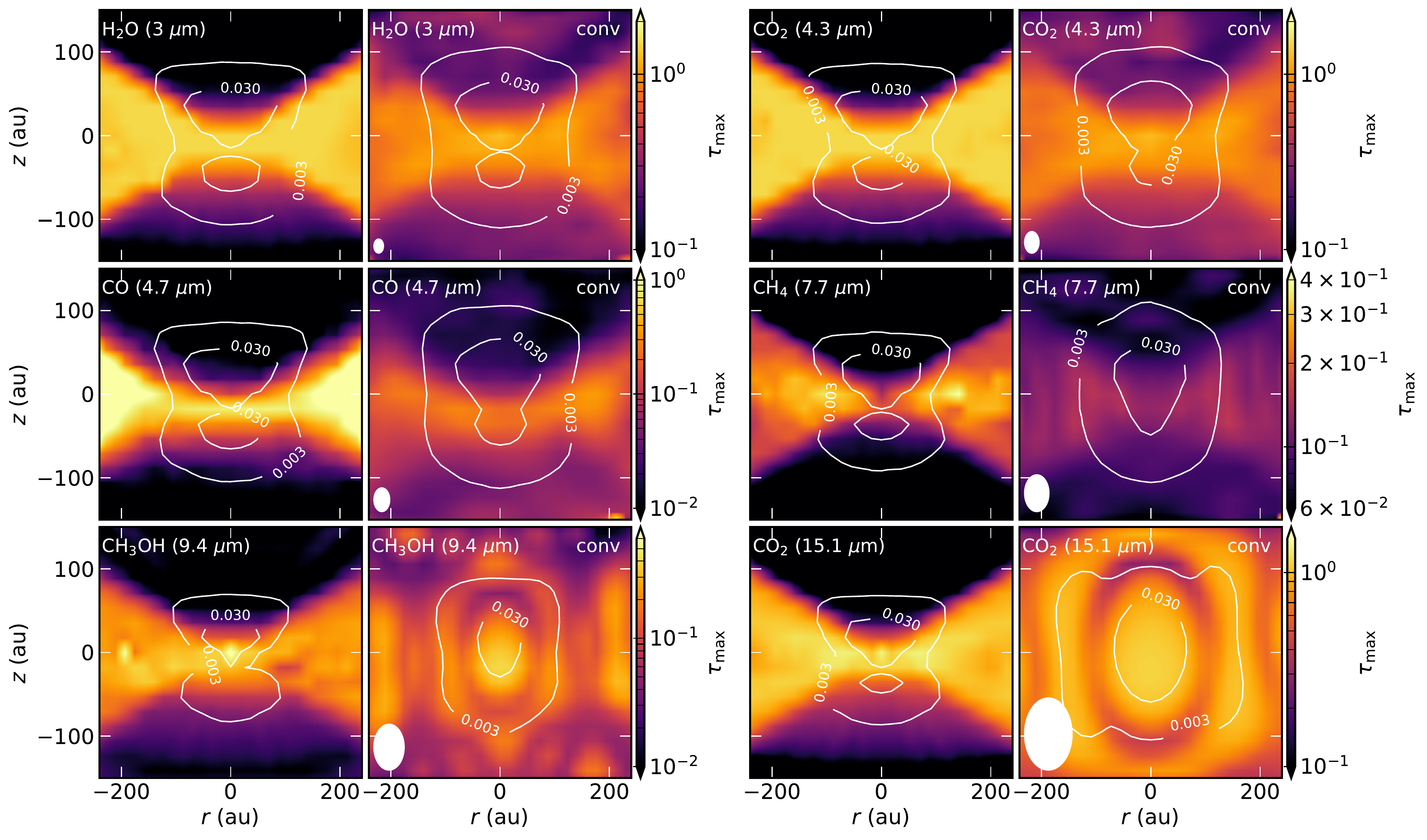}
    \caption{The expected peak optical depth for the 6 main ice absorption features as a function of disk radius and height towards HH~48~NE. 
    Left columns shows the optical depth in the model at similar pixel resolution as NIRSpec and MIRI. 
    Right columns show the same after convolving the model with an instrumental PSF of NIRSpec (<5~\micron) or MIRI (>5~\micron) respectively. 
    The FWHM of the PSF is shown in the lower left corner. 
    White contours show the local continuum at 0.003 and 0.030 mJy for reference. 
    30 $\mu$Jy is a typical 3$\sigma$ sensitivity of a JWST/ NIRSpec exposure of $\sim$1 hour. Note the different color scales for each ice feature.}
    \label{fig:peaktaupos}
\end{figure*}

In Fig. \ref{fig:fiducial_ice_sed}, we present the mid-infrared spectrum of the fiducial model, and compare it to the observations in the four Spitzer IRAC bands and the ``dust only" model.  
Adding ices to the model has a minor effect on the continuum with respect to the ``dust only" model, because of the addition of the ice scattering opacities (see also Fig. \ref{fig:opacities}).
The modeled spectrum is overall a good fit to the photometric data points in the considered spectral range.

\subsubsection{Integrated optical depths}
We fitted a local continuum around the main ice absorption features using a couple wavelength points just outside the (known ab initio) absorption features and a second order polynomial fit (see Fig. \ref{fig:fiducial_ice_sed}), and converted the spectrum to a logarithmic ice optical depth using
\begin{equation}
    \label{eq:optical depth}
    \tau = -\mathrm{ln}\left(\frac{F_{\nu}}{F_{\nu\mathrm{,cont}}}\right).
\end{equation}

The prominent \ce{H2O} ice feature at 3~\micron has a peak optical depth of 0.6 in the integrated spectrum (see Fig. \ref{fig:fiducial_ice_sed}). 
The feature has a minor contribution from \ce{CH4}, \ce{CH3OH} and \ce{NH3} in the model, compared to the \ce{H2O} O-H stretching mode.
The \ce{CO2} features at 4.26 and 15.2~\micron are also present in the spectrum, with similar strength as the 3~\micron \ce{H2O} ice feature. 
The feature at 15.2~\micron is double peaked, as expected for a pure \ce{CO2} ice \citep{Warren1986}. 
Adding a component of \ce{CO2} in polar mixtures will change the shape of this feature drastically, but this will be discussed in a forthcoming paper.
The \ce{CO} feature at 4.7~\micron has a peak optical depth of 0.07. 
Additional ice features in the spectrum are the \ce{CH4} feature at 7.7~\micron and the \ce{CH3OH} feature at 9.4~\micron. 
Those features have comparable strength to the CO feature.
The spectrum is dominated by photons that encountered multiple scattering events (compared with grey line in Fig. \ref{fig:fiducial_ice_sed}).
The ice features in the scattering dominated part of the spectrum (<10~\micron) are substantially weaker when only considering single scattering, since singly scattered photons will be scattered from a warm layer high up in the disk.

\subsubsection{Spatially resolved optical depths}
We determined the optical depth of the ice features as a function of disk radius and height at a similar pixel scale as that of the NIRSpec and MIRI instruments (Fig. \ref{fig:peaktaupos}).
We convolved the simulated observations with a NIRSpec or MIRI PSF, depending on the wavelength, and compare it to the original model (second and fourth column).
After convolving with an instrumental PSF, the vertical changes are smoothed out over a larger area, lowering the peak optical depths up to a factor of two.
The smallest resolvable scale at 4~\micron (see Fig. \ref{fig:peaktaupos}) is $\sim$20~au, while the resolvable scale at 15~\micron is $\sim$90~au. 
This means that at the MIRI longer wavelengths the disk will not be resolved in a lower and upper lobe. 

Fig. \ref{fig:peaktaupos} illustrates that the \ce{H2O} ice feature at 3~\micron is expected to be present throughout the disk. 
There is a clear variation in optical depth of the \ce{H2O} ice feature with the height in the disk, ranging from an optical depth of 1 at the midplane to 0.3 in the upper layers in the disk.
Note that the dark lane is offset from the midplane because of the 82.3$^\mathrm{o}$ inclination, so the strongest absorption features are found slightly below the central disk midplane.
In the dark lane, where the dust column density is the highest and temperatures are low, we trace a higher column density of ice, so naturally the optical depth of the water ice features is higher there than in the disk atmosphere.

Radial variations in peak optical depth are small in the model, even though the continuum changes over two orders of magnitude, the ice column density decreases rapidly with radius and the optical depth is not saturated.
This is in contrast to dark cloud ice observations, where in general the ice column density is probed directly, and illustrates the importance of radiative transfer in understanding edge-on disk ice absorption.
The small radial variations in optical depth show that most absorption happens in the outer disk regions that are crossed similarly by light from all lines of sight. 

The \ce{CO2} ice features at 4.3~\micron and 15.1~\micron are predicted to absorb in similar regions of the disk as the water ice (Fig. \ref{fig:peaktaupos}). 
For the ice features at wavelengths >8~\micron, it will be impossible to see vertical changes, due to the extent of the instrumental PSF.
CO is only available near the disk midplane, as the ice is vertically limited by the temperature structure in the disk (see Fig. \ref{fig:denstemp}).
The CO ice feature is strongest in the lower surface of the disk, but we note that the absorption in the upper surface of the disk contributes equally to the integrated spectrum due to the stronger continuum in the upper surface (see Fig. \ref{fig:peaktaupos_weighed}).
The vertical CO snowline is resolved in the disk upper surface, but the edge is smoothed out due to radiative transfer effects and PSF smoothing.

\begin{figure*}
    \centering
    \includegraphics[height = .94\textheight]{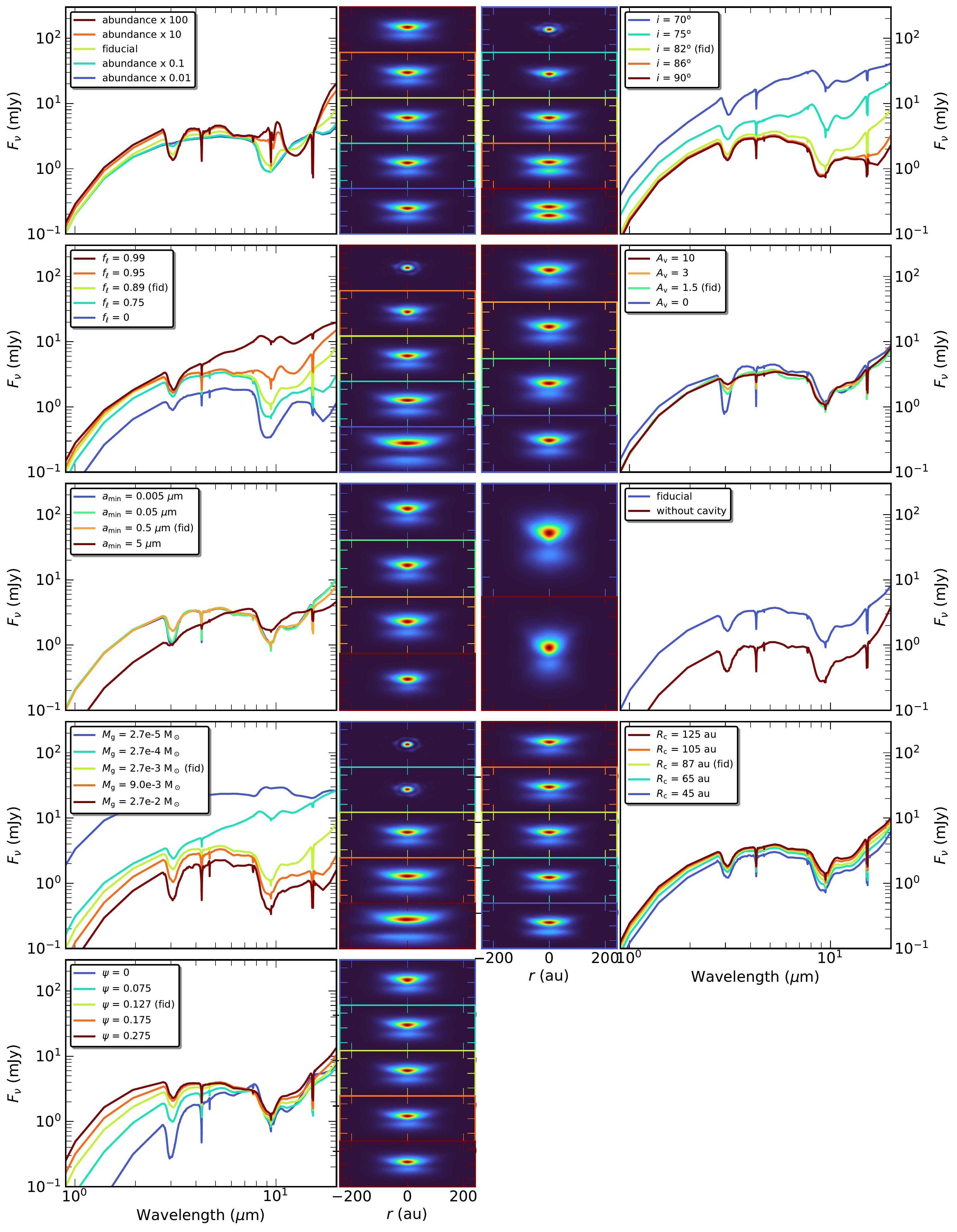}
    \caption{Large panels: sensitivity of the mid-infrared spectrum to various parameters of the model, stated in the legend. Small panels: The images show the appearence of the disk in scattered light at 4~\micron, convolved with the NIRSpec PSF. Panels are labeled with the same color as the legend and shows to what extent changes in these parameters will be visible in the NIRSpec images. The vertical ticks indicate $\pm$ 100~au.}
    \label{fig:icesed_comparison}
\end{figure*}

\subsection{Sensitivity to parameters}
\label{ssec:ice_sensitivity_to_params}
In Fig. \ref{fig:icesed_comparison}, we present the modeled spectra for a range of parameters compared to the fiducial model that best resembles the currently available continuum observations (see Paper I for details).
The evolution of the integrated peak optical depth with these parameters is presented in Fig. \ref{fig:total_dependence}. 
Note, however, that these peak optical depths are dominated by the disk regions with highest continuum values (see Fig. \ref{fig:peaktaupos_weighed}).
To make comparisons between models at different disk locations more universal, we have summed all emission in a radial slab at 7 different heights in the disk and show the optical depth in Figs. \ref{fig:deps1}, \ref{fig:deps2}, and \ref{fig:deps3}. 

\subsubsection{Sensitivity to abundance}
We varied the abundance of all molecules by two orders of magnitude from the fiducial model in 9 logarithmic steps, while keeping the total amount of mass the same.
Fig. \ref{fig:icesed_comparison} shows the integrated spectra and the spatially-resolved results are shown in Fig. \ref{fig:deps1}.
In a direct observation of ice against a background source instead of a scattering continuum, one would expect a peak optical depth that is directly proportional with abundance, as the ice column density is thought to scale linearly with optical depth.
In the limit of low ice abundances (less than ISM abundance) we recover the same trend for edge-on protoplanetary disks (see large panel in Fig. \ref{fig:total_dependence}). 
However, this slope flattens towards high abundances for all ice species, before reaching total saturation of the absorption feature (see Fig. \ref{fig:icesed_comparison}). 
This is likely a result of ``local saturation", meaning that the ice absorption features are saturated at the position in the disk where the ice absorbs the light, but that additional light from the same line of sight contributes to the total flux that is not saturated because it scattered through disk regions with less ice.
The result is that low ice abundances can be inferred with great accuracy, but for abundances >10x the ISM abundance can only be constrained with a lower limit.
The optical depth of \ce{CO2}, \ce{H2O}, and \ce{CH3OH} can reach higher values than \ce{CH4} and CO due to their larger freeze out region and therefore higher absolute absorption. 
The maximum optical depth can vary from source to source depending on the scattering continuum and source structure.


\subsubsection{Sensitivity to fraction of settled mass}
\label{ssec:Sensitivity to fraction of settled mass}
The amount of settling (expressed as mass fraction in large, settled grains: $f_\ell$, with corresponding fraction of material left in the atmosphere: 1-$f_\ell$) has a direct effect on the height of the scattering surface in the disk, defined as the height at which the disk becomes optically thick to stellar light at each wavelength. 
This in turn determines what part of the disk will absorb the scattered light from the star and the warm inner disk and thus the depth of the ice absorption features.
We changed the amount of settled dust mass in the model setup from 0 to 99\% in 5 steps (see Figs. \ref{fig:icesed_comparison}, \ref{fig:deps1}) to determine how this affects the ice spectrum.
Note that $f_\ell$ = 0 implies that there are no grains larger than 7~\micron in the disk, which means that the longer wavelengths for that model are missing an important emission contribution. 

Settling has a major impact on the scattered light images, and scattering continuum.  
Less settling means that the upper and lower emission surfaces of the disk extend further out in height, and the width of the dark lane increases. 
This is a direct result of the height of the scattering surface, which increases if the amount of small grains in the disk atmosphere increases.
Disks with $f_\ell$ >99\% do not fully shield the star at an inclination of 82.3\degr and the source appears as a point source.
Since the settling has a major effect on the resolved scattered light observations, the JWST observations provide tight constraints on the amount of small dust grains in the disk atmosphere.

The ice features are strongly affected by the amount of settling (see Fig. \ref{fig:total_dependence}). 
There are multiple effects at play at the same time, which result in a complicated dependence that is different for every ice feature.
The amount of settled dust mass changes the position of the scattering surface, as explained earlier, which affects the amount of cold material under the scattering surface.
This effect is visible in Fig. \ref{fig:total_dependence} as a general increase of the strength of the ice features towards less settled disks up to an order of magnitude due to an increased column density of cold material under the scattering surface.
However, for all molecules except CO the optical depth decreases again for $f_\ell$ $\gtrsim$0.9. 
This turnover is due to a directional effect where the light is scattered more through the disk surface rather than through the cold midplane. 
The same effect creates the increased width of the dark lane for a less settled disk.
Additionally, the upper layers of the disk become more and more optically thin, due the settling of material, which allows for more direct stellar light or singly scattered light that traces a smaller column of ice.
The \ce{h2o} feature at 3~\micron doesn't follow the trend up in opacity from $f_\ell$ = 0.99 to $f_\ell$ = 0.9, which is likely because \ce{h2o} has a high desorption temperature, thus is available throughout the disk.  Also, this feature at a relatively short wavelength is less affected by direct stellar light.  

\subsubsection{Sensitivity to inclination}
In previous works, it has been shown that the inclination can have a strong effect on the visibility of the ices in the disk \citep[e.g.,][]{Terada2017,Ballering2021}.
We varied the inclination of our model from 60$^{\mathrm{o}}$ to 90$^{\mathrm{o}}$ in 8 steps to determine the impact on the absorption features in HH~48~NE.

\begin{figure*}
    \centering
    \includegraphics[width = \textwidth]{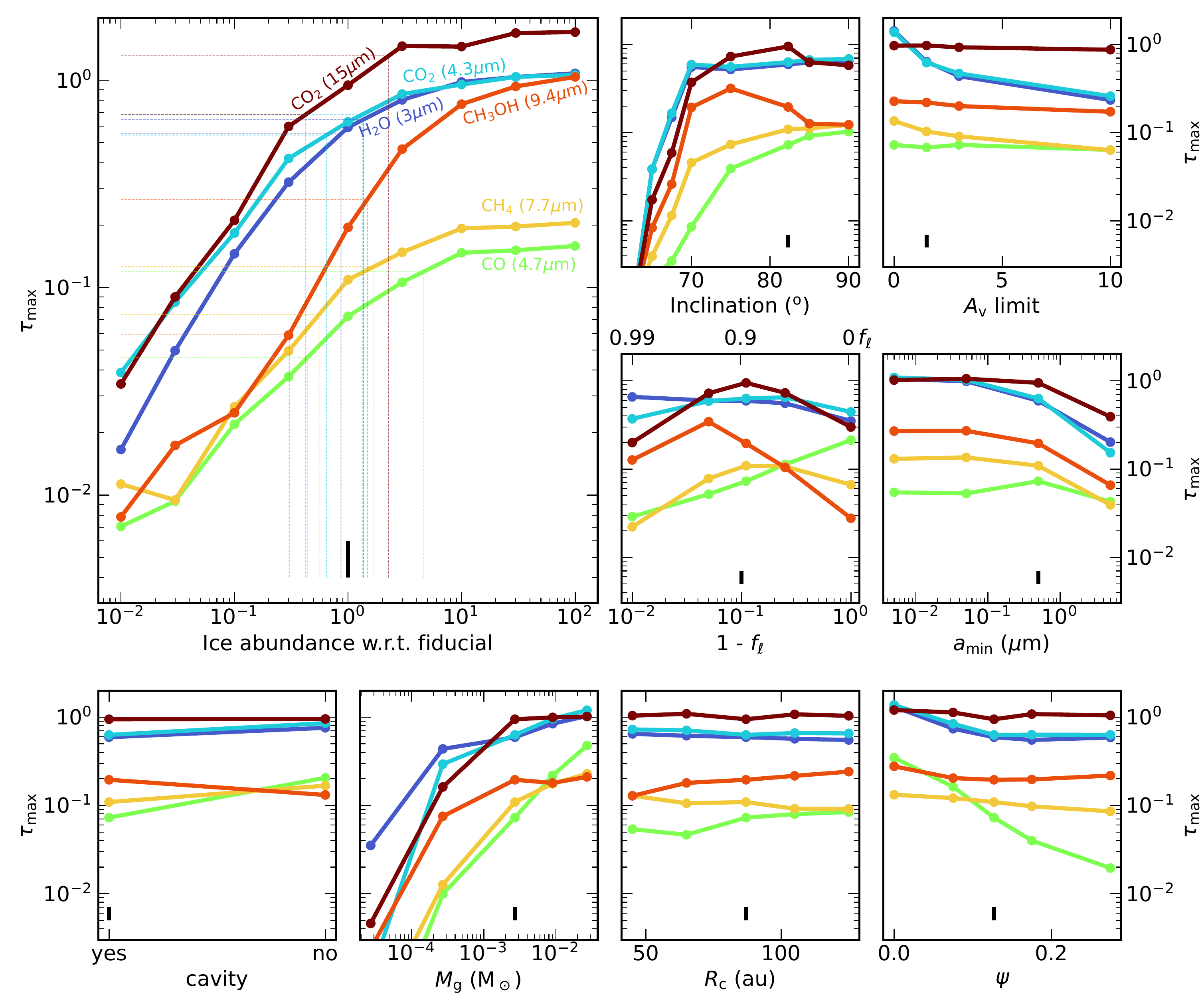}
    \caption{Sensitivity to various parameters of the maximum optical depth of the six main ice features. The black vertical line is the fiducial model. The dotted lines in the large panel indicate the uncertainty of the peak optical depth (horizontal) and corresponding abundance (vertical) when taking the uncertainty of all parameters into account. Varied parameters are: ice abundance, inclination, $A_{\rm v}$ limit, mass fraction of small grains expressed as 1-$f_\ell$, minimum grain size, existence of a cavity, disk mass and characteristic disk radius. A typical deep observation of JWST has a contrast ratio of 300 on the continuum (i.e., S/N = 300), which translates to a 3$\sigma$ peak optical depth of $\sim$10$^{-2}$.}
    \label{fig:total_dependence}
\end{figure*}

We show in Fig. \ref{fig:icesed_comparison} that the total spectrum of the disk changes over an order of magnitude, but the optical depth of the ice features is not very sensitive to the inclination for most ice species (see Fig. \ref{fig:total_dependence}), as long as the disk is sufficiently edge-on to block direct lines of sight to the star ($i$ >70\degr; see Fig. \ref{fig:fiducial_ice_sed}). 
This is consistent with previous modeling work that show little variation in ice optical depths >80\degr \citep{Pontoppidan2005}, although that model predicts significantly more absorption close to the critical inclination of 70\degr. 
This difference is likely a result of the differences between the model setups, as \citet{Pontoppidan2005} did not include any settling of large grains nor an $A_v$ threshold for UV photo-dissociation of ices which results in a larger amount of ice in the atmospheric layers traced in disks at an inclination of $\sim$70\degr compared to our model.
The sharp transition around $i$ $\sim$70\degr is confirmed observationally using all known water ice detections in protoplanetary disks \citep{Terada2017}.
The CO ice feature in the integrated spectrum increases by an order of magnitude from 70 to 90\degr. 
This occurs primarily because, at low inclination the scattered light continuum does not reach the coldest regions in the midplane of the disk along the line of sight, but will only probe the ice in layers higher up in the disk.

At lower inclinations the upper surface of the disk absorbs significantly less in the \ce{H_2O}, \ce{CO2} and CO features than the lower surface (see Fig. \ref{fig:deps1}).
The asymmetry between absorption depth in the upper half versus the lower half could thus be a direct probe of the disk inclination.
For low inclinations (<75$^{\mathrm{o}}$), the peak of the absorption features shifts to longer wavelengths, which is seen in all major ice features (see Fig. \ref{fig:deps1}).
At those inclinations the ice scattering opacity plays an important role and causes shifts in the peak position and changes in the absorption profiles. 
In extreme cases, this results in absorption features with a wing in emission in \ce{CO2} \citep{dartois2022}.

\subsubsection{Sensitivity to $A_{\rm v}$ limit}
As mentioned in Sect. \ref{sec:model_description} we added a physically motivated $A_{\rm v}$ limit to the height in the disk where ice is allowed to form on the dust grains (see Fig. \ref{fig:denstemp}).
Above this limit ices will be easily desorbed by non-thermal processes under the intense irradiation of high-energy stellar photons \citep{boogert2015,Arabhavi2022}.
We varied the $A_{\rm v}$ limit from the fiducial value of 1.5 to 0, 3, and 10~mag.
Without an $A_{\rm v}$ limit the optical depth of \ce{H2O} and \ce{CO2} remains fairly constant over the height of the disk, even up to 90~au from the disk midplane (Fig. \ref{fig:deps2}).
With the $A_{\rm v}$ limit in place, there is less absorption of \ce{H2O} and \ce{CO2} (Fig. \ref{fig:total_dependence}), and a larger variation with height (see Fig. \ref{fig:deps2}). 
The $A_{\rm v}$ limit has no effect on the CO absorption, as the regions where the disk is cold enough for CO to freeze out are already lower than the $A_{\rm v}$ limit (see Fig. \ref{fig:denstemp}) and the effect on the continuum is minor (Fig. \ref{fig:icesed_comparison}).
The ice features are not sensitive to small variations in the limit that is used.
The peak of the absorption features moves with increasing $A_{\rm v}$ limit to longer wavelengths.
This is the case for the \ce{H2O}, the \ce{CO2} and the silicate feature (see Sect. \ref{ssec:silicate_feature} for more details).

\subsubsection{Sensitivity to grain size}
To determine the sensitivity of the ice features to the grain size distribution, we changed the minimum grain size from the fiducial model of 0.5~\micron by extending it to smaller grains. 
The absorption features of most ice features, except CO, are slightly stronger if we include small grains up to the canonical value of ISM grains, 0.005~\micron, but there is no significant difference between a minimum grain size of 0.005~\micron and 0.05~\micron.
Ices on large grains are partly invisible, as shown in previous studies \citep[e.g.][]{dartois2022}.
Including more small grains will therefore boost the fraction of observed ice, hence the optical depth.
Any addition of grains smaller than 1~\micron is enough to dominate the absorption, and including more small dust grains does not change the ice optical depths significantly. 
If only grains larger than 5~\micron are used, the optical depth changes significantly by a factor of $\sim$5 (see Fig. \ref{fig:total_dependence}).

\subsubsection{Sensitivity to a cavity}
To determine the effect of a cavity on the spectrum, we ran the same model as the fiducial model, but without a dust depletion inside the cavity.
The integrated continuum is suppressed by a factor of 3-4 (see Fig. \ref{fig:icesed_comparison}), but the cavity has negligible effect on the \ce{H2O} and \ce{CO2} ice features (see Fig. \ref{fig:total_dependence}). 
This indicates that most absorption happens in the outer regions of the disk and that the cavity does not have a significant effect on the region where the warm dust emission is scattered in the disk, as long as there is warm material in the inner disk.
The CO feature at 4.67~\micron does change by a factor of 3, which is considerable given that a similar boost could result from an elevated abundance by an order of magnitude. 

\subsubsection{Sensitivity to total mass}
The disk mass is varied over 4 orders of magnitude from 3$\times10^{-5}$ - 3$\times10^{-2}$ M$_\odot$, while keeping the gas-to-dust ratio constant.
We would like to note that a lower disk mass therefore decreases the ice column densities, similar to the ice abundance grid, but also has a significant effect on the scattering continuum (see Fig. \ref{fig:icesed_comparison}).
Ice absorption features are sensitive to order of magnitude changes, but only the features at wavelengths <5~\micron change significantly if the mass increases an order of magnitude.
By changing the disk mass the temperature structure changes, as well as the column density of material and the height of the emitting layer. 
The latter two dominate in the impact on the ice features. 
This results in a weaker change of the features at longer wavelengths, partly due to local saturation explained in Sect. \ref{ssec:Sensitivity to fraction of settled mass}. 
This is especially true for the methanol feature at 9.4~\micron and the \ce{CO2} feature at 15.2~\micron. 
The \ce{CO2} feature at 4.27~\micron is much more affected than the \ce{CO2} feature at 15.2~\micron, which could serve as a direct probe of the disk mass. 
Note however that the feature at 15.2~\micron can change considerably in shape if the \ce{CO2} is in different ice matrices.

\subsubsection{Sensitivity to radial size}
The physical size of the disk can play a major role in setting its temperature structure. 
We varied one of the parameters that determines the radial extent of the disk, $R_\mathrm{c}$, between 45 and 125~au.
Although the effect on the continuum is significant (Fig. \ref{fig:icesed_comparison}), the ice optical depths do not change significantly (Fig. \ref{fig:total_dependence}). 
We would like to note that for models with a small $R_\mathrm{c}$, the 55~au cavity pushes material outwards to sustain the same disk mass as the fiducial model. 
Increasingly small disks without a cavity are expected to show variations due to changes in the area below the snow surfaces.

\subsubsection{Sensitivity to flaring index}
 We varied $\psi$ from a flat disk ($\psi$ = 0) to an extremely flared disk ($\psi$ = 0.275) in 5 steps $\psi$ $\in$ [0, 0.075, 0.127, 0.175, 0.275].
The flaring has only a minor effect on the ice features in the MIRI wavelength regime, but the optical depth of the ice features <5~\micron increases for a flatter disk.
The flaring of the disk ($\psi$) changes the temperature structure in the disk and directly affects the angle of the scattering surface towards the midplane.
For disks with less flaring, this angle is smaller, thus more light will reach the ice-rich regions of the disk.
This effect is the strongest for the CO ice feature at 4.7~\micron, because CO ice is only located near the midplane. 

\subsection{Silicate feature}
\label{ssec:silicate_feature}
Not only the ice absorption features, but also the shape and strength of the silicate feature at 10~\micron are affected by the physical parameters of the disk model (see Figs. \ref{fig:deps1}, \ref{fig:deps3}). 
In the fiducial model, the silicate feature is only seen in absorption, unlike the typical emission feature seen in disks with an inclination <70\degr \citep{Olofsson2009}. 
Absorption of the silicate feature arises if the scattering surface is high up in the disk and the light has to pass through a large column of cold dust in the outer disk.

The silicate feature is very sensitive to the inclination. 
For inclinations <75$^{\mathrm{o}}$, an emission component is seen in addition to the silicates in absorption and at even lower inclinations the silicate feature appears only in emission. 
Very settled disks do also show a silicate feature in emission, as the scattering surface is low in the disk, which means that the warm dust above the scattering surface significantly adds emission to the spectrum. 
These changes also have effect on the determined \ce{CH3OH} optical depths (see Fig. \ref{fig:total_dependence}).
The silicate feature varies in strength with varying ice abundance, and is even seen in emission if the grains are coated thick enough with ice.

\begin{figure}[ht]
    \centering
    \includegraphics[width = \linewidth]{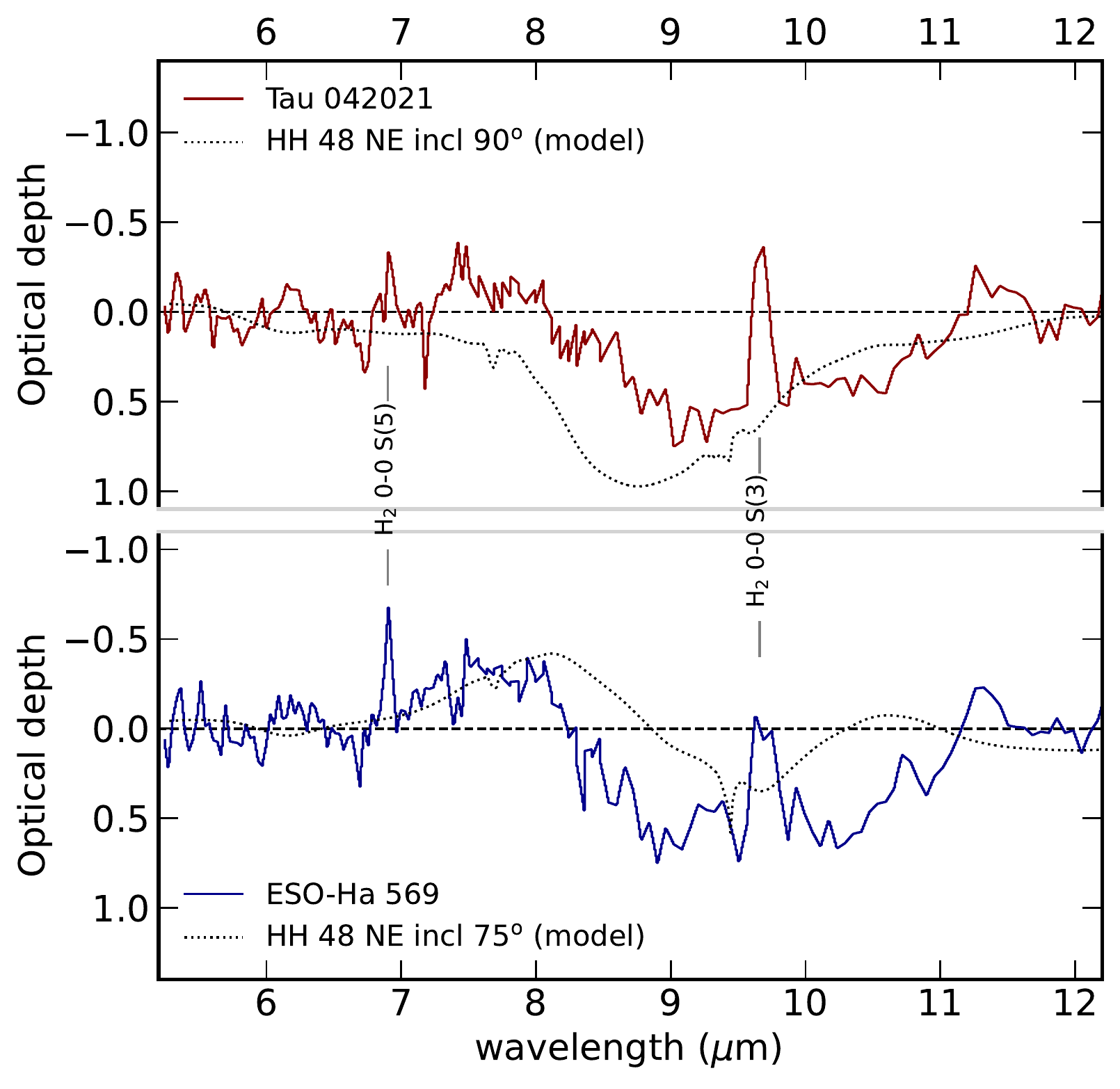}
    \caption{Optical depth of the silicate feature in Spitzer spectra for Tau~042021 and ESO-Ha~569. The black lines show the optical depth of the silicate features in a model inclined at 90\degr and 75\degr. The shape of the silicate feature is clearly a function of inclination and disk mass.}
    \label{fig:spitzer_comparison}
\end{figure}

Changes in the shape of the silicate absorption feature are observed in Spitzer IRS spectra towards edge-on protoplanetary disks.
For comparison, we retrieved low resolution spectra from the CASSIS archive \citep{Lebouteiller2011_cassisLRS} of the edge-on disks Tau~042021 and ESO-Ha~569 (see Fig. \ref{fig:spitzer_comparison}).
The optical depth of the silicate feature is determined with a linear fit using data points between (5,6) and (11,14) \micron and Eq. \ref{eq:optical depth}.
The silicate feature in ESO-Ha~569 has a clear emission component around 8~\micron, while the feature in Tau 042021 has only a component in absorption. 
Without detailed modeling of the sources, we overplot the model for HH~48~NE at an inclination of 90\degr and 75\degr.
The inclination of these sources can roughly be inferred from their silicate feature alone, as ESO-Ha~569 is inclined at $\sim$83\degr \citep{wolff2017}, while Tau~042021 is inclined at 88\degr \citep{Villenave2020}. 
The differences in inclinations are exaggerated, because of the differences in mass and settling of the two disks and HH~48~NE. 
An estimate of the dust mass is necessary to use this approach in observations, for example from millimeter continuum observations.
The shape and strength of the silicate feature can be crucial to break parameter degeneracies and help interpreting the ice observations.

\section{Discussion}
\label{sec:discussion}
The high sensitivity and resolution of JWST will provide spatially resolved measurements of ices in Class II protoplanetary disks. 
This observational data can only be properly interpreted with detailed modeling.
This is especially true for edge-on disks, since the ice absorption features are highly dependent on the physical source structure and are not easy to interpret. 
We have explored the effect of various parameters on the ice optical depth, that can be directly compared to observations to interpret the results.

\subsection{Constraining ice abundances}
We have carefully investigated what the impact of various physical disk parameters is on the optical depth of the main ice features in protoplanetary disks. 
We find that the molecular abundance and disk mass have the strongest effect on the observed ice features, which results primarily from a reduction of the absolute column density of ice.
The optical depth increases monotonically with the abundance for all ice features, but the exact shape is dependent on source structure and wavelength.
Two other important parameters are the inclination and the amount of grain growth and settling.
These parameters can change the optical depths in the same order of magnitude as the chemical abundance. 
It is hence crucial to have a well constrained physical disk model to constrain ice abundances from observations.

This work is the first to convolve the model with an instrumental PSF to see if we can resolve vertical changes in ice abundance.
HH~48~NE should be resolved sufficiently within the NIRSpec range to constrain vertical abundance gradients due to temperature, irradiation, and chemistry. 
At wavelengths >7~\micron, the vertical dependencies are no longer resolved, due to the increasing size of the PSF.
The vertical extent of the CO and \ce{CO2} snowline could therefore be observable in the derived observed optical depths, although the transition will be smooth since the elevation of the snow surface increases radially outwards.
The PSF sidelobes can induce significant artefact ice feature signatures throughout the disk, even in regions where direct raytracing of the model without convolution of a PSF shows no sign of ices (e.g., at 100~au in Fig. \ref{fig:peaktaupos}).
Since the radial dependence of the ice optical depth is largely insensitive to the disk geometry and most absorption happens in the outer regions of the disk, it will be difficult to infer radial snowlines or abundance variations in radial direction.
This is due to the fact that all photons originate from central regions in the disk, which means that all photons that are scattered in our direction have encountered a wide range of radii no matter what line of sight they appear to come from.
We would like to note that RADMC-3D uses long characteristics radiative transfer for the image reconstruction, which in general is more expensive but gives the most accurate results. We tested another radiative transfer approach combining both long characteristics and short characteristics radiative transfer, \texttt{Hyperion} \citep{Robitaille2011}, which gives identical results (see Appendix \ref{app:comparison_hyperion} for more details).

We have shown that the optical depth of the ice features increases monotonically with abundance, but an accurate estimate of the source inclination, amount of settling and disk mass is required to determine the conversion factor. 
The maximum optical depth that can be reached in our models is $\sim$1, even when going to very high ice abundances and column densities.
This illustrates that the ice features may be locally saturated in the disk, but that at least a third of the light is scattered through the ice-free upper layers of the disk, before being scattered in our direction from a layer deeper down in the disk.
The ice optical depths that we find in our model are relatively high compared to earlier work by \citet{Ballering2021} and \citet{Arabhavi2022}.
Contrary to these works, we expect to find CO in the integrated NIRSpec spectra and deeper \ce{CO2} ice features. 
\citet{Ballering2021} used a standard model with a high disk mass and complex gas-grain chemistry in the same radiative transfer code \texttt{RADMC-3D}.
Their continuum for an inclination of 85\degr is similar to our spectrum, except that their silicate feature is in emission even at an inclination of 90\degr. 
The water ice feature in their spectrum is of similar strength as in our model, but the other species compared to water ice are much weaker.
The final abundances in their inheritance model are very similar to our input abundance and are fairly constant with height and radius. 
The differences between their expected strength of the ice absorption features and this work are therefore largely a result of the geometry of the fiducial disk model.
Observed optical depths towards the class I disk in \citet{Pontoppidan2005} are much deeper than our model, but they show that most ices are located in the envelope surrounding the disk and only 50\% of \ce{H2O} and \ce{CO2} and none of the \ce{CO} ice are located in the disk.
Since HH~48~NE is a more evolved disk without thick envelope, absorption from cloud ice does not play a significant role.

In Sect. \ref{ssec:ice_sensitivity_to_params}, we show that global ice abundances can be inferred from spatially integrated observations if settling, inclination and disk mass are known. 
In the case of HH~48~NE, these three parameters are well constrained \citep[see Paper I;][]{PaperI}. 
Treating the uncertainties as independent from each other, we can derive a first order estimate of the uncertainty on measured ice abundances from the parameter variation in Fig. \ref{fig:total_dependence}. 
We determined the uncertainty on the optical depth expected from the 1$\sigma$ uncertainty interval via interpolation for each of the considered parameters.
Combining the uncertainties using $\sigma_{\rm tot}^2 = \sum_{\rm p}\sigma_{\rm p}^2$, we determine the uncertainty on the molecular abundance for each ice feature (see the vertical lines in Fig. \ref{fig:total_dependence} top left panel). 
The uncertainty on the ice abundances is typically a factor of 2 and only the CO abundance is uncertain to a factor of 3.

\subsection{Detectability of edge-on disks}
Highly inclined disks (i>$70^\mathrm{o}$) are statistically less often detected than their less inclined counterparts \citep{luhman2008,rodgerslee2014}. 
From this parameter study we identify several potential explanations for this trend.

First, edge-on disks are often selected based on their SEDs, and are then confirmed with scattered light imaging or resolved mm continuum observations.
Edge-on disks typically have a high infrared excess and a characteristic dip in the mid-infrared due to the silicate feature in absorption.
The spectra of these disks can look similar to transition disks, that show a comparable dip in the mid-infrared due to a lack of warm material in the inner disk regions \citep{VanderMarel2022_HRtransition_disks}, or to Class I sources that have less stellar contribution to their SED as well.
We have shown that edge-on disks with a high level of settling or with an inclination of <85$^\mathrm{o}$ may instead show a silicate feature in emission, which means that these objects may not be recognized as edge-on disks until they are observed at high angular resolution. 
Complete sample studies at high angular resolution with ALMA or scattered light observatories are hence necessary to increase the fraction of detected edge-on disks. 
Unfortunately, very high angular resolution is necessary to determine high inclined sources directly from the observations \citep{Villenave2020} and observed samples of disks in scattered light are far from complete.

Second, radial substructure might be important to detect edge-on disks. 
The HH~48~NE model without cavity has a factor of 3-4 lower scattered light continuum, which makes the disk harder to detect (see Fig. \ref{fig:icesed_comparison}).
With a cavity the mid-infrared is boosted due to structural changes in the disk density profile.

Third, in many instances we find that the disk disappears due to a lack of contrast between the star and the disk (see Fig. \ref{fig:icesed_comparison}). 
This is especially true for disks with a high level of settling, a low inclination or a low disk mass.
This results in an observation bias, as we see that the average observed disk mass for edge-on disks \citep[>25.6 M$_\oplus$][]{Villenave2020} is high compared to the mean dust mass distribution of the full Taurus and Lupus disk populations \citep[15 M$_\oplus$][]{Andrews2013,Ansdell2016}, especially when taking into account that edge-on disks are much more likely to be optically thick \citep{Villenave2020}.
Without sufficient amounts of material in the upper layers of the disk to block direct lines of sight to the star, it will be impossible to detect these disks in scattered light without a coronagraph \citep[see also][]{wolff2017}. 
The recent work by \citet{Angelo2023} comes to similar conclusions using a simulated population of edge-on protoplanetary disks. 

\section{Conclusions}
\label{sec:conclusion}
We have set up a disk model using the radiative transfer code \texttt{RADMC-3D} that includes an-isotropic scattering and opacities of the six most important ice species.
The sensitivity of the optical depth of the main ice features to various physical parameters is explored to see how well chemical abundances can be constrained in the disk. 
We can conclude the following:

\begin{itemize}
\item The main ice species, \ce{H2O} and \ce{CO2}, are likely detectable in JWST data of edge-on protoplanetary disks. 
Weaker features like those from CO, \ce{CH4}, \ce{CH3OH}, and \ce{NH3} require a high inclination and low amount of settling to be observed.
\item The optical depth of the ice features in protoplanetary disks changes monotonically, but non-linearly with ice abundance, and is saturated at an optical depth $\lesssim$1 due to the effect of local saturation.
Detailed modeling of the source structure is necessary to infer chemical effects at play in the disk.
Abundances in the HH~48~NE disk can be constrained within a factor of three with respect to the dust content, taking the current uncertainty on the disk geometry into account.
\item Vertical ice abundance gradients are detectable in the NIRSpec range (\ce{H2O}, \ce{CO2}, and CO), but resolved optical depths are not linearly related to the ice column density along the line of sight. 
For wavelengths longer than 7~\micron, resolved maps will be dominated by PSF residuals due to large angular scale of the PSF.
\item The shape of the silicate feature and the comparison between the two \ce{CO2} bands, at 4.3 and 15.1 \micron respectively, could help to break the degeneracy between total solid mass and ice-to-rock ratio.
\item The low detection rate of edge-on disks can be explained by a combination of silicate features appearing in emission at inclinations <85$^\mathrm{o}$, radial substructure that is necessary to boost the mid-infrared continuum and the need for a large amount of dust high up in the disk to block direct lines of sight towards the star.
\end{itemize}

We have shown that JWST will be able to observe spatially resolved ice features in protoplanetary disks. 
In depth modeling of the icy chemistry, disk geometry and radiative transfer is crucial to interpret forthcoming observations of the JWST ERS program IceAge and edge-on disks in general.

\begin{acknowledgements}
We would like to thank the anonymous referee for suggestions that improved the manuscript.
We also thank Ewine van Dishoeck for useful discussions and constructive comments on the manuscript.
Astrochemistry in Leiden is supported by the Netherlands Research School for Astronomy (NOVA), by funding from the European Research Council (ERC) under the European Union’s Horizon 2020 research and innovation programme (grant agreement No. 101019751 MOLDISK).
M.N.D. acknowledges the Swiss National Science Foundation (SNSF) Ambizione grant no. 180079, the Center for Space and Habitability (CSH) Fellowship, and the IAU Gruber Foundation Fellowship.

This work makes use of the following software: \texttt{numpy} \citep{numpy}, \texttt{matplotlib} \citep{matplotlib}, \texttt{astropy} \citep{astropy2013,astropy2018}, \texttt{OpTool} \citep{Dominik2021optool}, \texttt{RADMC-3D} \citep{Dullemond2012radmc3d}, \texttt{Hyperion} \citep{Robitaille2011}, \texttt{WebbPSF} \citep{webbpsf2012}.
\end{acknowledgements}

\bibliographystyle{aa}
\bibliography{refs.bib}

\begin{appendix}
\section{Comparison between \texttt{RADMC-3D}  and \texttt{Hyperion}}
\label{app:comparison_hyperion}
\begin{figure}[ht]
    \centering
    \includegraphics[width = \linewidth]{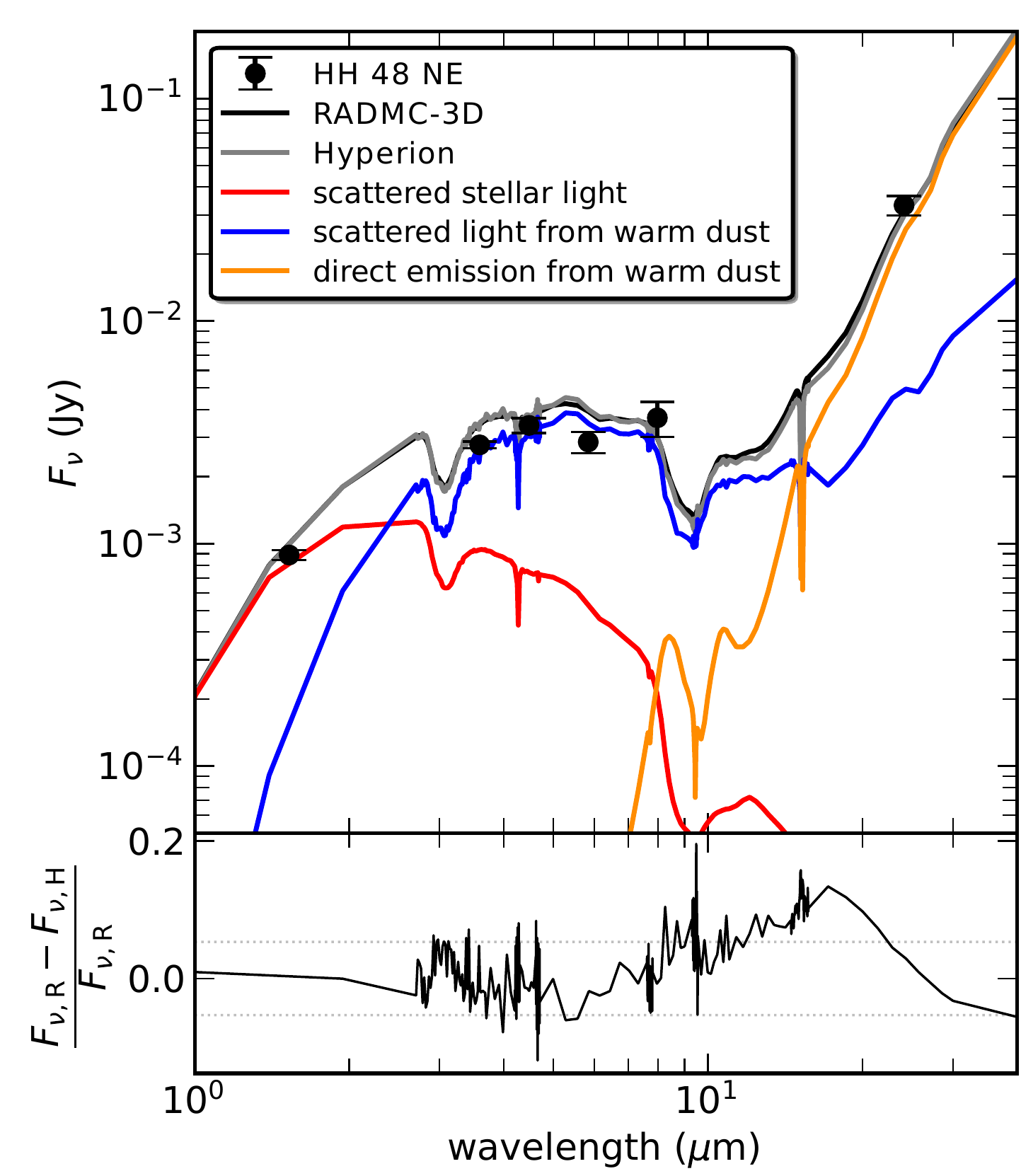}
    \caption{Comparison between raytraced spectra by \texttt{RADMC-3D}  and \texttt{Hyperion}. The top panel shows the SED for both \texttt{Hyperion}  and \texttt{RADMC-3D}. The colored lines show the contribution to the total spectrum of scattered stellar light (red), scattered light from warm dust (blue) and direct light from warm dust (orange). The direct stellar light component is negligible en therefore not shown in this plot. The bottom panel shows the normalized difference between the two total output of the two codes. The dashed lines indicates the standard deviation of the difference, which is 5\%.}
    \label{fig:hyperion_comparison}
\end{figure}

As a validation of the suitability of \texttt{RADMC-3D} for full anisotropic scattering in edge-on disk models, we ran the best fitting model once with the radiative transfer code \texttt{Hyperion}  \citep{Robitaille2011}. 
\texttt{Hyperion} treats scattering internally differently, tracing each individual photon package from source to destination. 
The benefit of this is that multiple inclinations can be raytraced at once, and that the final spectrum can be dissected in 4 components: direct stellar light, scattered stellar light, direct dust emission, and scattered light originated from warm dust (see Fig. \ref{fig:hyperion_comparison}). 
The downside is that the code is very demanding in terms of computing time and computer RAM (>500~Gb; hence only possible on high-performance clusters).
The relative difference between the final output of the two codes is less than 20\% everywhere in the spectrum, does not increase in the ice absorption features, and has a standard deviation of 5\%.
With this comparison we demonstrate that the output can be trusted independent of the radiative transfer code that is used.

In the spectral region where the important ice absorption features are located, the continuum is dominated by the dust scattering by half an order of magnitude over direct source scattering at short wavelengths and dust emission at long wavelengths.  

\section{Additional material}
\begin{figure*}
    \centering
    \includegraphics[width = \textwidth]{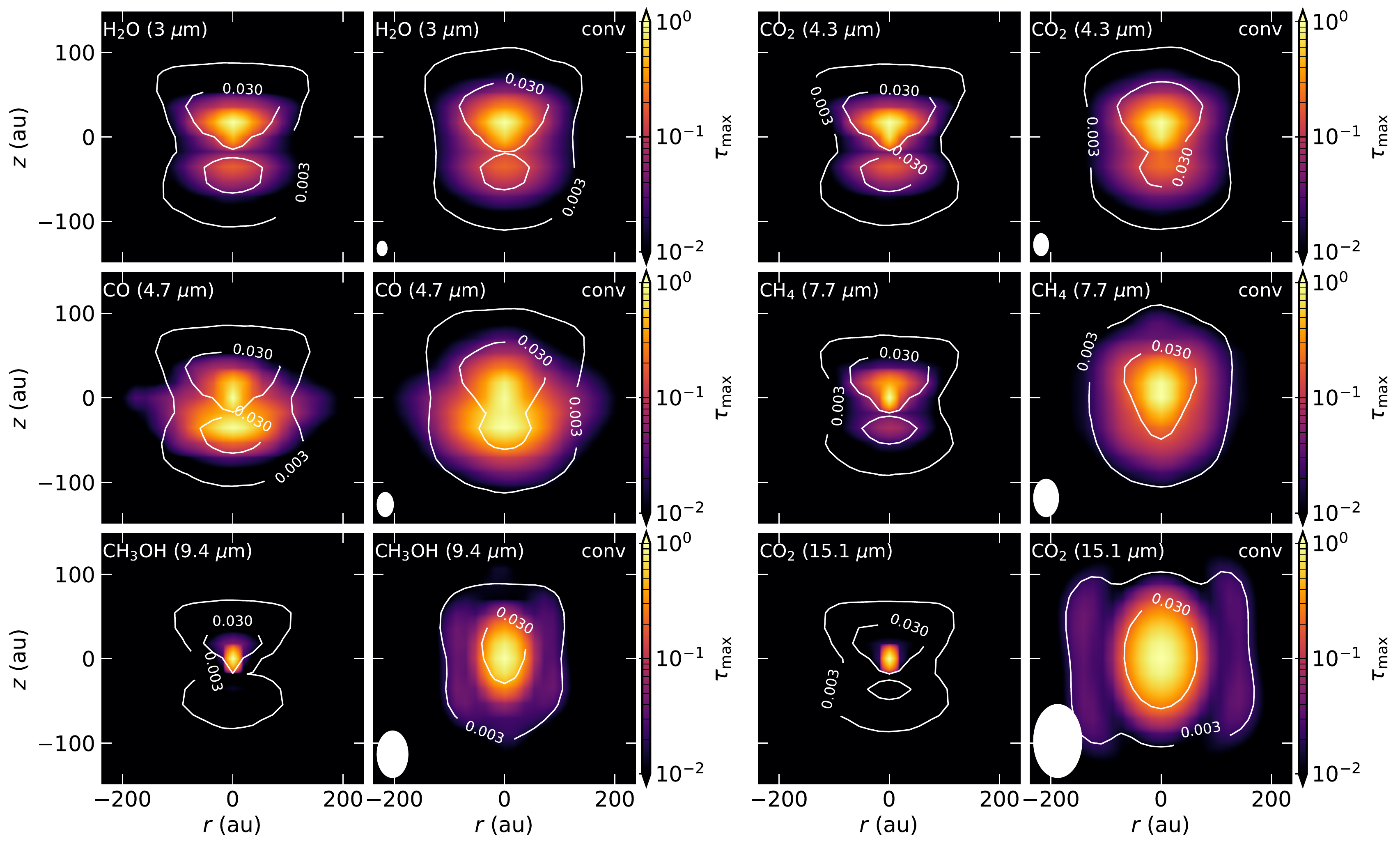}
    \caption{Same as Fig. \ref{fig:peaktaupos}, but weighted with the continuum emission and normalized to the peak contribution. This figure illustrates what part of the disk contributes mostly to the disk averaged spectrum or the radial slabs presented in Figs \ref{fig:deps1}, \ref{fig:deps2}, and \ref{fig:deps3}.}
    \label{fig:peaktaupos_weighed}
\end{figure*}

In Fig. \ref{fig:peaktaupos_weighed} we present the spatially resolved optical depth of the ice features, similar to Fig. \ref{fig:peaktaupos}, but weighted with the continuum emission.
The maps show how much the different disk regions contribute to the disk integrated spectrum.
In all absorption features, only a few pixels determine the optical depth of the ice in the disk integrated spectrum.
Spatially resolved observations are necessary to trace any variations in abundance, directly.

\begin{figure*}
    \centering
    \includegraphics[width=0.95\textwidth]{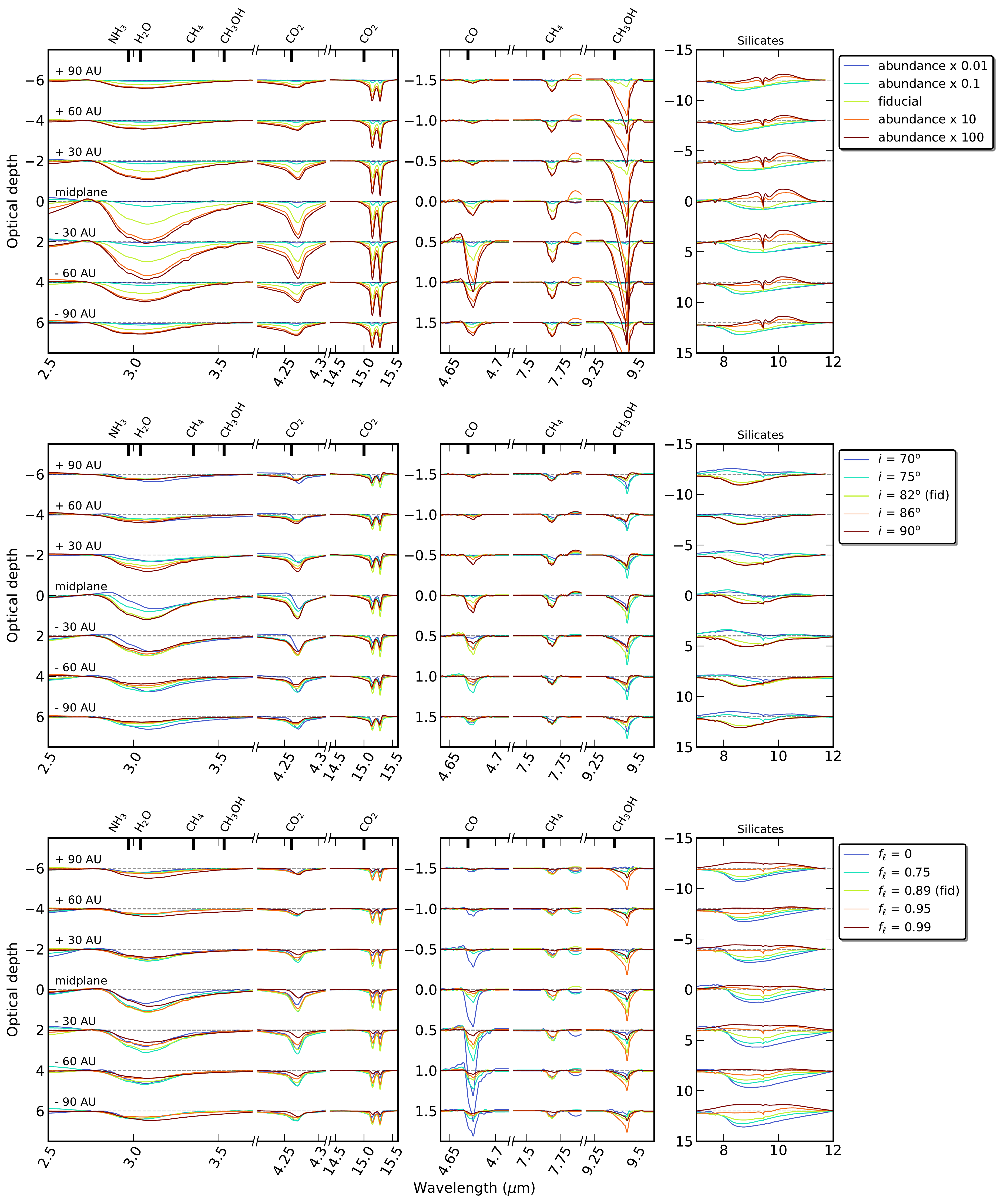}
    \caption{Dependence of the ice optical depth on the ice abundance, the inclination and the fraction of settled large grains. The modeled optical depth of the main ice features and the silicate feature are shown at 7 different vertical positions in the disk, convolved with an instrumental PSF. The resolution in height is 30~au, which corresponds to two pixels in NIRSpec/MIRI data.}
    \label{fig:deps1}
\end{figure*}

\begin{figure*}
    \centering
    \includegraphics[width=.95\textwidth]{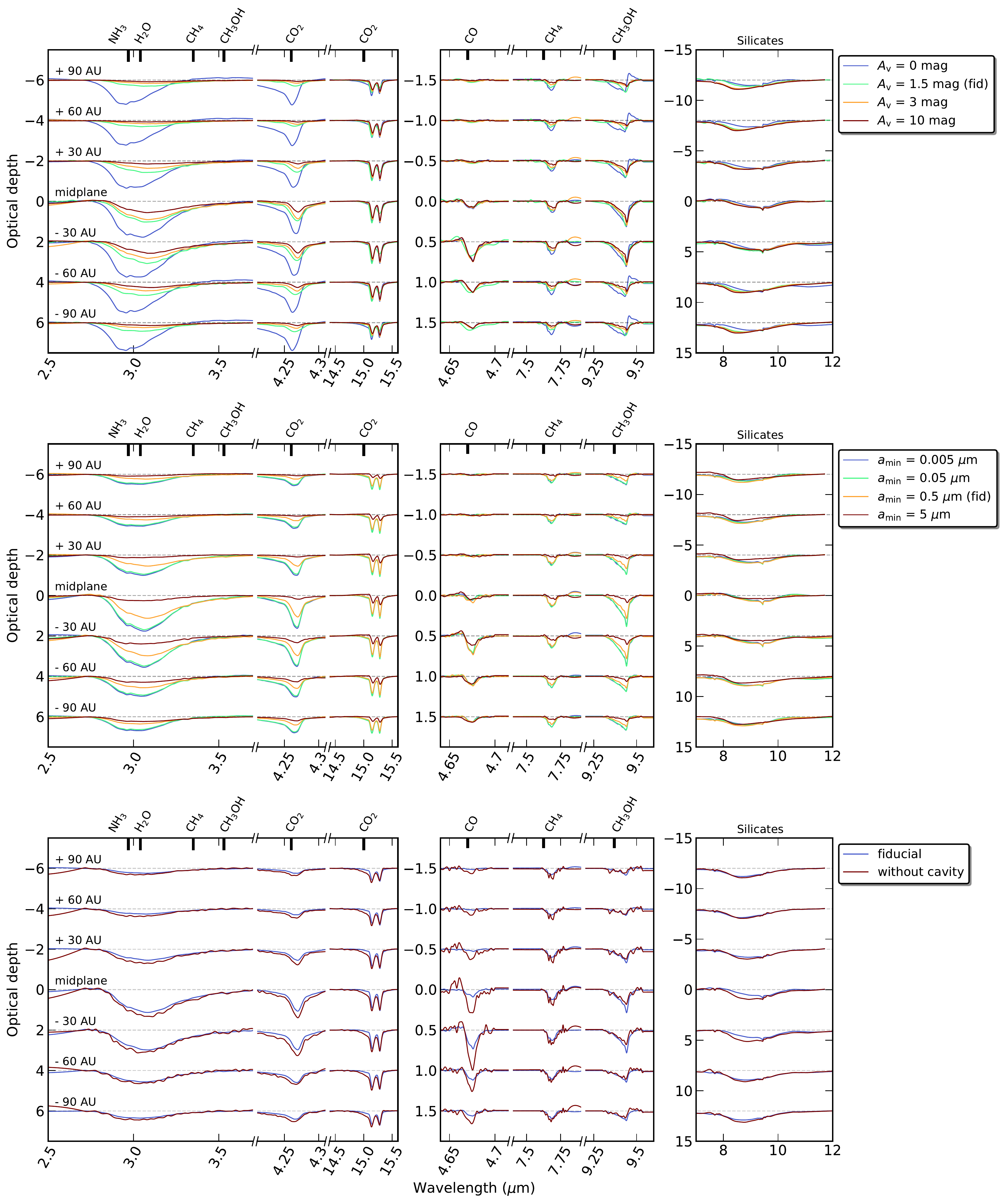}
    \caption{Same as Fig. \ref{fig:deps1}, but instead varying the $A_{\rm v}$ limit, the maximum grain size as well as with and without a dust cavity.}
    \label{fig:deps2}
\end{figure*}

\begin{figure*}
    \centering
    \includegraphics[width=.95\textwidth]{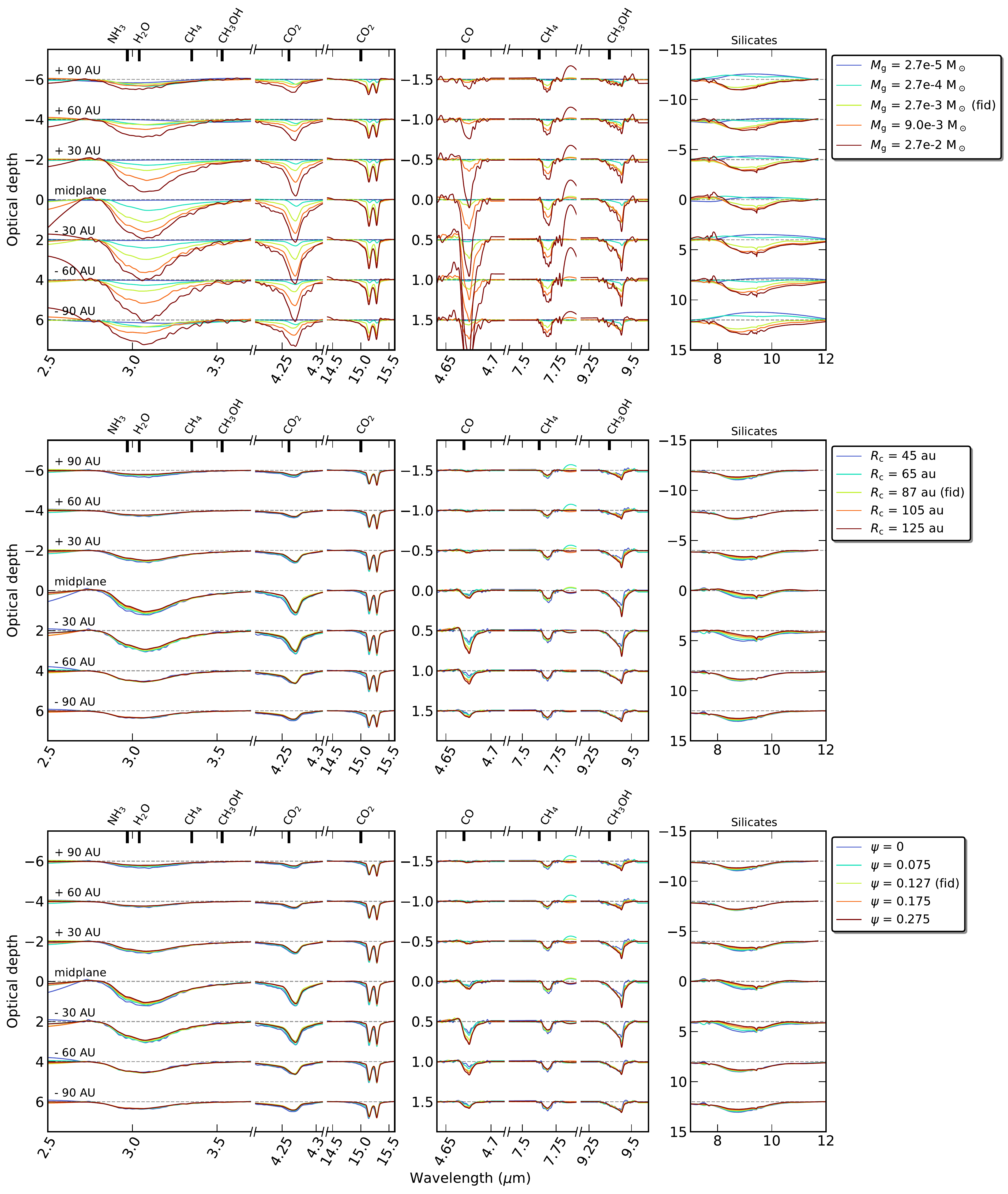}
    \caption{Same as Fig. \ref{fig:deps1}, but instead varying the total disk mass and the spatial scale of the disk as a function of $R_{\rm c}$.}
    \label{fig:deps3}
\end{figure*}

In Figs. \ref{fig:deps1}, \ref{fig:deps2}, and \ref{fig:deps3}, we present the vertically resolved optical depth of the main ice absorption features as function of the same parameters discussed in Sect. \ref{ssec:ice_sensitivity_to_params}


\end{appendix}
\end{document}